\documentclass[aps,showpacs,twocolumn]{revtex4-2}
\usepackage[utf8]{inputenc}
\usepackage{amsfonts}
\usepackage{amssymb}
\usepackage{amsmath}
\usepackage{graphicx}
\usepackage{appendix}
\usepackage{hyperref}
\usepackage{color}

\hypersetup{hypertex=true, colorlinks=true, linkcolor=blue, urlcolor=blue, citecolor=blue}

\begin{document}

\title{Hilbert space fragmentation in quantum Ising systems induced by side
coupling}
\author{E. S. Ma}
\author{Z. Song}
\email{songtc@nankai.edu.cn}
\affiliation{School of Physics, Nankai University, Tianjin 300071, China}
\begin{abstract}
We study Hilbert space fragmentation and quantum scars in quantum spin
systems with Ising interactions. The system consists of two sets of quantum
spins, A and B. As the parent system, set A is an Ising model on arbitrary
lattices with a transverse field, while set B comprises free spins that are
coupled to set A. We show that the Hilbert space is fragmented into
exponentially many decoupled sectors when the transverse field and the side
coupling strength are at resonance. As examples, several typical systems
with quantum scars are studied analytically. Numerical simulations of
probability distribution of entanglement entropy for finite-size chains,
square and triangular lattices are performed using the Monte Carlo method.
The results show that Hilbert space fragmentation and the corresponding
quantum scars become pronounced when the system approaches resonance.
\end{abstract}

\maketitle

%\affiliation{College of Physics and Materials Science, Tianjin Normal University, Tianjin
%300387, China}

%\affiliation{School of Physics, Nankai University, Tianjin 300071, China}

%\email{zhangxz@tjnu.edu.cn}

\section{Introduction}

Quantum information processing must overcome not only the well-known
challenge of environmentally-induced decoherence but also the subtler threat
of thermalization. The eigenstate thermalization hypothesis (ETH) not only
explains thermalization in isolated systems within the framework of quantum
mechanics \cite%
{Deutsch_Quantum,srednicki1994chaos,d2016quantum,borgonovi2016quantum,gogolin2016equilibration,serbyn2021quantum}%
, but also seems to pose challenges for quantum simulation and quantum
information tasks. Fortunately, evidence shows that the ETH can be violated
in some situations \cite%
{nandkishore2015many,abanin2019colloquium,bernien2017probing,choi2019emergent,turner2018,pai2019localization,kwan2025minimal,feng2022hilbert,zhao2020quantum,turner2021correspondence,mukherjee2020collapse,van2020quantum,bluvstein2021controlling,surace2021exact,yang2025constructing}%
. Most eigenstates still follow the ETH, yet non-thermal behavior can be
observed when the system is prepared in some special initial states. A
promising mechanism for the anomalous thermalization is the Hilbert space
fragmentation. Beyond the many-body localization (MBL) and integrablility of
a system, fragmentation is a third mechanism can prevent thermalization,
challenging the ETH.\textbf{\ }As promising mechanism for the anomalous
thermalization, it originates from intrinsic kinetic constraints \cite%
{schecter2019weak,yang2020hilbert,moudgalya2022hilbert,li2023hilbert,francica2023hilbert,nicolau2023flat}%
, which fragment the Hilbert space into dynamically isolated subspaces,
thereby rendering some states inaccessible and preventing full
thermalization. Constrained models, such as the PXP model \cite%
{lesanovsky2012interacting,turner2018}, constrained spin chains \cite%
{lingenfelter2024exact}, and dipole-conserving hopping models \cite%
{sala2020ergodicity}, were the first to exhibit fragmented dynamics, which
is indicative of Hilbert space fragmentation. In systems with fragmented
Hilbert spaces, certain subspaces may contain special eigenstates that are
the quantum many-body scars (QMBS) \cite{shiraishi2017,
moudgalya2018,moudgalya20182,khemani2019,ho2019,shibata2020,mcclarty2020,richter2022,jeyaretnam2021,turner2018,turner20182,shiraishi2019,lin2019,choi2019emergent,khemani2020,dooley2020,dooley2021, he2026Hilbert}%
. In the ref. \cite{he2026Hilbert}, the connection of Hilbert space
fragmentation and QMBS is investigated in hardcore Bose and Fermi Hubbard
models in the framework of the restricted spectrum generating algebra \cite%
{moudgalya2020eta}. These non-thermal states are typically embedded within
the bulk spectrum of the system and span a subspace in which initial states
fail to thermalize and instead exhibit periodic behavior. In practice, the
kinetic constraint is usually not imposed naturally, but induced from the
particle-particle interactions. Consequently, the corresponding interaction
strength determines the degree of the Hilbert space fragmentation, which
then influences the formation of the scar.

\begin{figure}[tbh]
\centering
\includegraphics[width=0.4\textwidth]{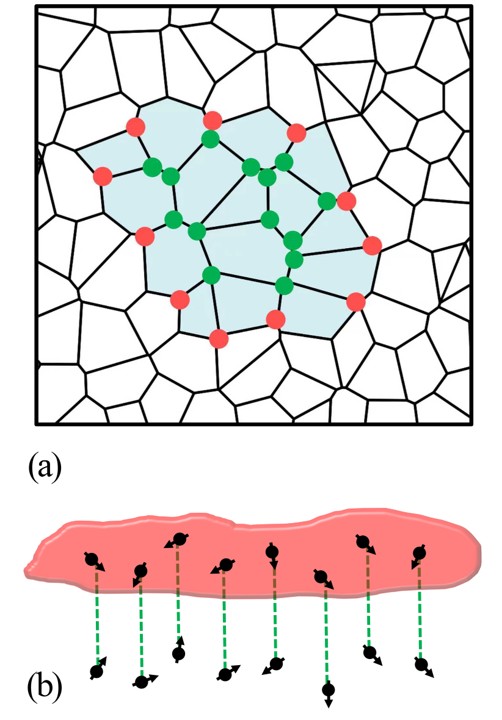}
\caption{Schematic illustration of the main idea of this work. (a) We
consider an Ising model with a transverse field on an arbitrary lattice.
When the local transverse field at a site vanishes, the spin state at this
site is pinned. When a set of such pinned spins (red dots) forms the
boundary of a region (green dots), the kinetic constraint for the domain
walls results in Hilbert-space fragmentation. (b) The structure of a system
with side couplings. It consists of two sets of quantum spins, an upper set
and a lower set. The red shading represents the Ising interaction within the
upper set of spins, where each spin also couples to its counterpart in the
lower set.}
\label{fig1}
\end{figure}

In this work, we propose a class of quantum spin systems that exhibit
Hilbert space fragmentation. The key feature of this scheme is its structure
of the system, which consists of a parent Hamiltonian and an auxiliary
coupling counterpart. The parent Hamiltonian is a standard Ising model with
a transverse field on an arbitrary lattice, where each spin couples via an
Ising-type interaction to its non-interacting counterpart. Importantly, this
side coupling can effectively cancel the transverse field and pin the spin
state when resonance conditions are met. This results in Hilbert space
fragmentation when a set of pinned spins forms a closed boundary that
confines domain walls. Fig. \ref{fig1} provides a schematic illustration of
this scheme. To investigate the efficiency of the fragmentation, numerical
simulations of probability distribution of entanglement entropy\ for
finite-size chain, square, and triangular lattices are performed using the
Monte Carlo method. The results show that the Hilbert space fragmentation
becomes pronounced when the system approaches resonance. In addition, a
typical subspace of quantum scars in a uniform bipartite lattice is studied
analytically and numerically for both on-resonance and off-resonance cases.
Our findings provide concrete examples that are beneficial for understanding
the mechanism of Hilbert space fragmentation.

The structure of this paper is as follows. In Sec. \ref{Model and HSF}, we
introduce the model Hamiltonian and show that the Hilbert space fragments in
the resonant case. In Sec. \ref{Entanglement entropy at resonance}, we
investigate the effects of Hilbert space fragmentation by analyzing the
entanglement entropy of eigenstates for several lattice geometries. In Sec. %
\ref{Quantum scars}, we provide examples to demonstrate that such systems
possess quantum scars. Finally, in Sec. \ref{conclusions}, we provide a
summary and discussion.

\begin{figure*}[tbh]
\centering
\includegraphics[width=1.0\textwidth]{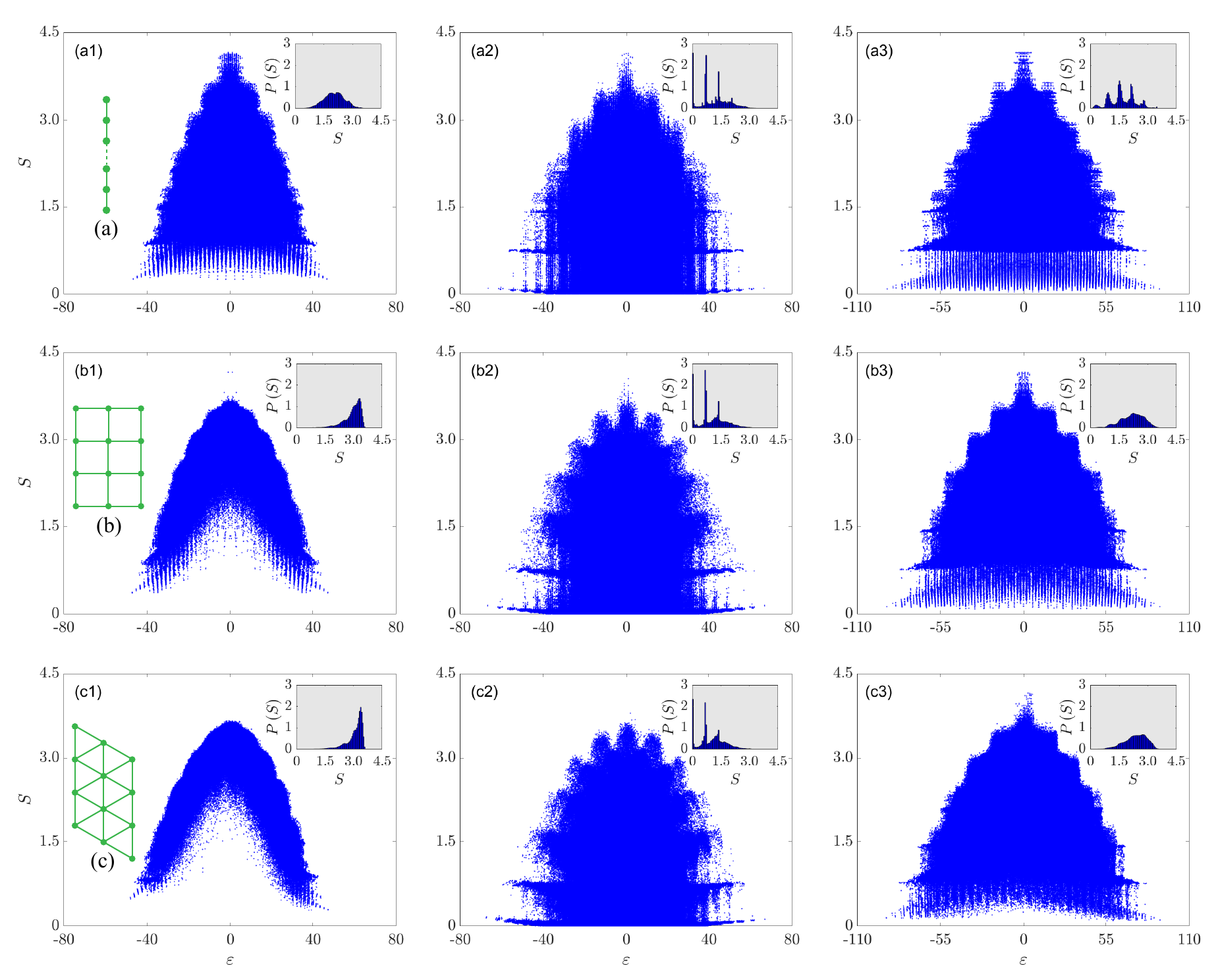}
\caption{The entanglement entropy $S$ for the eigenstates of the Hamiltonian
in Eq. (\protect\ref{H}) for three geometries of lattice A shown in the
insets [(a) an open chain, (b) a square lattice, and (c) a triangular
lattice] and the corresponding probability density distribution $P(S)$. $S$
denotes the entanglement entropy between the odd sites and even sites of an
eigenstate with eigenenergy $\protect\varepsilon $. The parameters $\protect%
\kappa _{j}$ in Eq. (\protect\ref{H}) and $g_{j}$ in Eq. (\protect\ref{H0})
are uniform, and $\protect\kappa _{j}=3, J=1$ are fixed. (a1), (a2), and
(a3) correspond to $g_{j}=1$, $g_{j}=3$, and $g_{j}=5$, respectively. (b)
and (c) have the same parameter settings as (a). The size of the system is
taken as $N=12$, and the numerical simulations are performed by utilizing
the Monte Carlo method. Specifically, we randomly select 1000 eigenstates of
lattice B in the basis of $\protect\sigma ^{x}$, with each constructing an
invariant subspace of the complete Hamiltonian. These results indicate that
the proportion of zero entropy significantly increases for the resonant case
where $g=\protect\kappa $ and that the distribution of entropy is affected
by the geometry of the lattice. }
\label{fig2}
\end{figure*}

\section{Model and Hilbert space fragmentation}

\label{Model and HSF}

The system we consider consists of two sets of quantum spins, A and B. Set A
is an Ising model with a transverse field, while set B comprises free spins
that are coupled to set A. Considering a $2N$-site system, the Hamiltonian
takes the form%
\begin{equation}
H=H_{0}+\sum_{j=1}^{N}\kappa _{j}\sigma _{a,j}^{x}\sigma _{b,j}^{x},
\label{H}
\end{equation}%
where the first term%
\begin{equation}
H_{0}=\sum_{\left\langle i,j\right\rangle }^{N}J_{ij}\sigma _{a,i}^{z}\sigma
_{a,j}^{z}+\sum_{j=1}^{N}g_{j}\sigma _{a,j}^{x},  \label{H0}
\end{equation}%
describes the transverse-field Ising model for set A, and the second term
represents the coupling between sets A and B. Here $\sigma _{\lambda
,j}^{\alpha }$\ ($\lambda =a,b;\alpha =x,y,z$) denotes the Pauli operators
on site $j$ of set $\lambda $. The parameters $g_{j}$, $J_{ij}$, and $\kappa
_{j}$\ represent the transverse field, the intraset coupling strength, and
the interset coupling strength, respectively. Lattice A is an arbitrary
lattice. The spins in lattice B do not interact with each other; instead,
they are only coupled to their counterpart spins in lattice A. Independently
of the parameter values $g_{j}$, $J_{ij}$, and $\kappa _{j}$, we note that 
\begin{equation}
\left[ H,\sigma _{b,j}^{x}\right] =0,  \label{symmetry}
\end{equation}%
which indicates that there exists a large number of local, mutually
independent conserved quantities. In the following, we will show that the
Hilbert space breaks into many disconnected sectors under certain conditions.

For simplicity, we start with a 1D Ising system and assume uniform
parameters, $g_{j}=g$, $J_{ij}=J$, and $\kappa _{j}=\kappa $. Considering
the case in which the two spins at ($b,j$)\ and ($b,j+r$)\ in the set B are
in the state 
\begin{equation}
\left\vert \varphi \right\rangle _{\text{\textrm{B}}}=\left\vert \leftarrow
\right\rangle _{b,j}\left\vert \leftarrow \right\rangle _{b,j+r}\prod_{l\neq
j,j+r}\left\vert \rightarrow \right\rangle _{b,l},
\end{equation}%
\ there exists a set of eigenstates, which can be written in the form%
\begin{equation}
\left\vert \psi \right\rangle =\left\vert \varphi \right\rangle _{\text{%
\textrm{A}}}\left\vert \varphi \right\rangle _{\text{\textrm{B}}},
\end{equation}%
with $\sigma _{b,l}^{x}\left\vert \rightarrow \right\rangle
_{b,l}=\left\vert \rightarrow \right\rangle _{b,l}$ and $\sigma
_{b,l}^{x}\left\vert \leftarrow \right\rangle _{b,l}=-\left\vert \leftarrow
\right\rangle _{b,l}$. Here the state $\left\vert \varphi \right\rangle _{%
\text{\textrm{A}}}$\ is an eigenstate of the Hamiltonian of lattice A
\begin{equation}
H_{\text{\textrm{A}}}=H_{0}-\kappa \left( \sigma _{b,j}^{x}+\sigma
_{b,j+r}^{x}\right) +\sum_{l\neq j,j+r}\kappa \sigma _{a,l}^{x}.
\end{equation}%
Importantly, when $\kappa =g$, the Hamiltonian%
\begin{equation}
H_{\text{\textrm{A}}}=J\sum_{l}\sigma _{a,l}^{z}\sigma _{a,l+1}^{z}+2\kappa
\sum_{l\neq j,j+r}\sigma _{a,l}^{x},
\end{equation}%
describes an Ising chain, in which the transverse field on the two spins at (%
$a,j$)\ and ($a,j+r$) in the set A are removed. Therefore, the eigenstate $%
\left\vert \varphi \right\rangle _{\text{\textrm{A}}}$\ can be written in a
form in which the spin states at sites $j$ and $j+r$ are pinned at the
eigenstates of $\sigma _{a,j}^{z}$ and $\sigma _{a,j+r}^{z}$, respectively.
This corresponds to an effective breaking the Ising chain at sites $j$\ and $%
j+r$. It results in several independent subspaces in the Hilbert space.
These two sites can be regarded as effective blockades of moving domain
walls. In this sense, the resonance condition $\kappa =g$ imposes a kinetic
constraint that allows spin flips at site ($a,l$) only when the spin at ($b,l
$) is in state $\left\vert \rightarrow \right\rangle _{b,l}$. Consequently,
many basis states are mutually inaccessible, and the Hilbert space fragments
into disconnected sectors. Accordingly, the original Hamiltonian at
resonance can be decomposed into $2^{N}$ independent Ising Hamiltonians of
type $H_{\text{\textrm{A}}}$. Furthemore, the Hilbert subspace that belongs
to each $H_{\text{\textrm{A}}}$\ breaks into many sectors. It appears that
the symmetry given in Eq. (\ref{symmetry}) plays a crucial role. However,
the fragmentation cannot be solely attributed to the symmetry, because an
additional resonance condition is required. Unlike ordinary symmetry
sectors, this fragmentation creates an exponentially many small, dynamically
isolated subspaces.

Now we turn to a more general case of the Hamiltonian $H$ in Eq. (\ref{H}).
Due to the symmetry given in Eq. (\ref{symmetry}), we have%
\begin{equation}
\left\vert \varphi \right\rangle _{\text{\textrm{B}}}=\prod_{j=1}^{N}\left%
\vert \lambda _{j}\right\rangle _{b,j},
\end{equation}%
where $\left\vert \lambda _{j}\right\rangle $ is an eigenstate of $\sigma
_{b,j}^{x}$, satisfying $\sigma _{b,j}^{x}\left\vert \lambda
_{j}\right\rangle =\lambda _{j}\left\vert \lambda _{j}\right\rangle $ with $%
\lambda _{j}=\pm 1$. Similarly, under the condition $g_{j}=\kappa _{j}$ for
all $j$, the original Hamiltonian can be decomposed into $2^{N}$ independent
Ising Hamiltonians of the form%
\begin{equation}
H_{\text{\textrm{A}}}(\left\{ \lambda _{j}\right\} )=\sum_{\left\langle
i,j\right\rangle }^{N}J_{ij}\sigma _{a,i}^{z}\sigma
_{a,j}^{z}+\sum_{j=1}^{N}\kappa _{j}\left( 1+\lambda _{j}\right) \sigma
_{a,j}^{x},
\end{equation}%
where $\left\{ \lambda _{j}\right\} =\{\lambda _{1},\lambda _{2},...,\lambda
_{N}\}$ is a configuration of $N$ binary numbers with $\lambda _{j}=\pm 1$.
The geometry of the lattice A is arbitrary. We note that when $\lambda
_{l}=-1$, the spin at ($a,l$) is pinned for any eigenstate of $H_{\text{%
\textrm{A}}}(\left\{ \lambda _{j}\right\} )$.\ The following case must exist
among all possible $\left\{ \lambda _{j}\right\} $. Consider $n$ sites with $%
\lambda _{j}=-1$ that form a boundary of a region, separating the lattice
into two sublattices. Any eigenstate of $H_{\text{\textrm{A}}}(\left\{
\lambda _{j}\right\} )$\ is expressed as the tensor product of states on two
sublattices.\ Then the Hilbert subspace of this $H_{\text{\textrm{A}}}$\ is
fragmented into many sectors. The structure of lattice is illustrated in
Fig. (\ref{fig1}).

An intuitive picture of such fragmentation is presented from the Hamiltonian 
$H_{\text{\textrm{A}}}(\left\{ \lambda _{j}\right\} )$ on a chain system. As
is well established that the Majorana representation of this Hamiltonian is
a $2N$-site chain with alternating hopping strengths $iJ_{ij}/2$\ and $%
i\kappa _{j}\left( 1+\lambda _{j}\right)/2$, respectively. We note that when 
$\lambda _{l}=-1$, the chain is disconnected at the corresponding dimer,
resulting in Hilbert space fragmentation. For a higher-dimensional system,
the corresponding Majorana lattice still exists. In this situation, although
the Majorana representation does not benefit the solution of the system, the
disconnections still remain valid.

\begin{figure*}[tbh]
\centering
\includegraphics[width=0.8\textwidth]{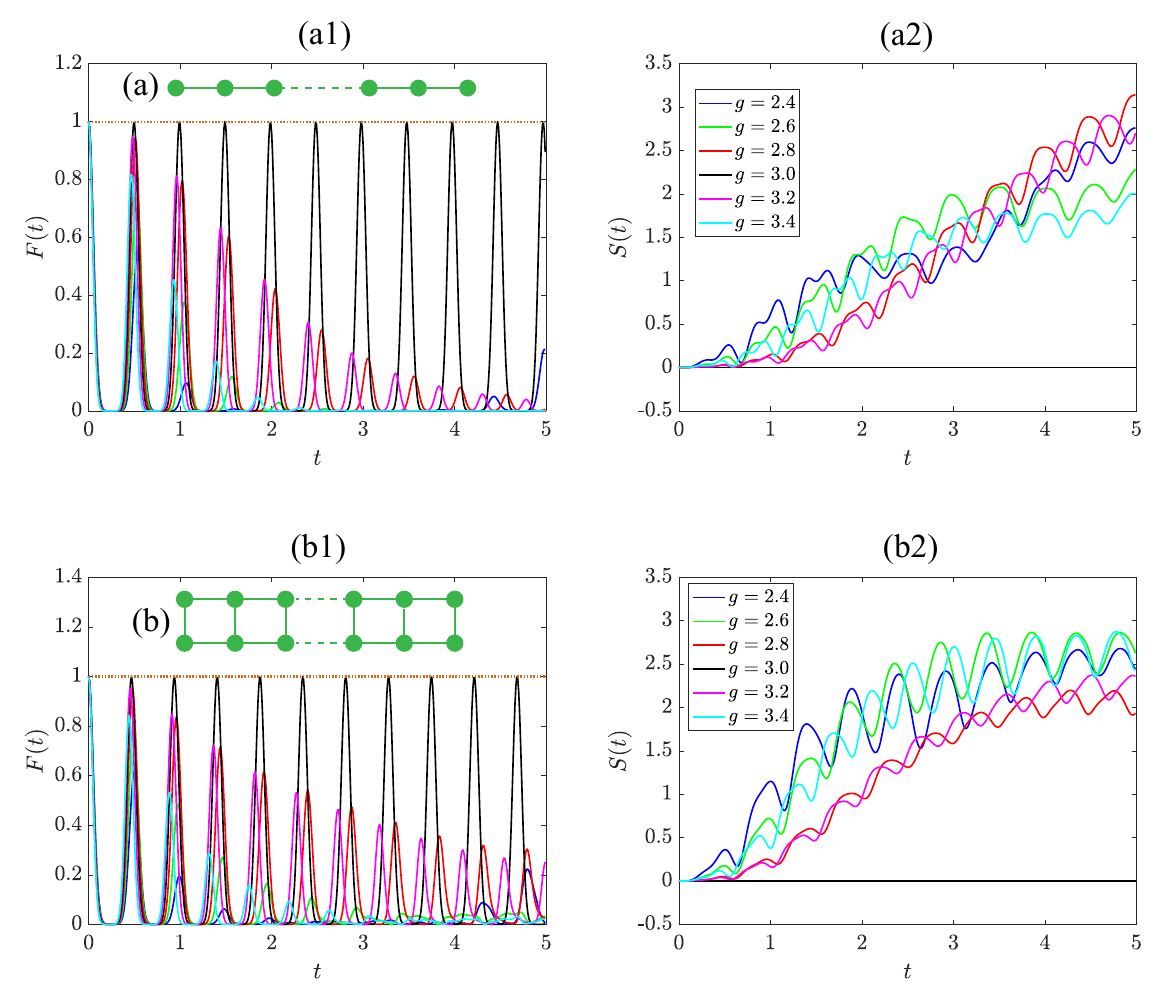}
\caption{Panels (a1) and (b1) display the dynamic fidelity defined in Eq. (%
\protect\ref{fidelity}) for two structures of lattice A with periodic boundary conditions, illustrated in the insets of 
(a) and (b). Panels (a2) and (b2) present
the corresponding bipartite entanglement entropy between odd sites and even sites. The
Hamiltonian is given by Eq. (\protect\ref{HA}), with lattice B prepared in the state specified by Eq. (\protect\ref{psiB}). Other parameters are set to $J=1$, $\protect\kappa=3$, and the total number of sites is 12. For the resonant case
where $g=\protect\kappa$, the fidelity demonstrates perfect periodic
oscillations while the entanglement remains at zero---both hallmark signatures of quantum many-body scars.}
\label{fig3}
\end{figure*}

\section{Entanglement entropy at resonance}

\label{Entanglement entropy at resonance}

In this section, we demonstrate our results in several typical finite-size
systems by numerical simulation. It is well known that the Hilbert space
fragmentation results in the existence of special eigenstates which exhibit
a small, sub-thermal amount of entanglement \cite%
{sala2020ergodicity,moudgalya2022quantum}. According to our results, the
resonance condition plays a crucial role in the Hilbert space fragmentation.

Quantum entanglement serves as a probe of thermalization and its breakdown
for a given eigenstate. Presumably, the number of eigenstates with vanishing
entanglement entropy increases significantly when the system approaches the
resonance. The complete set of eigenstates for a given system is obtained by
the numerical diagonalization of the Hamiltonian $H_{\text{\textrm{A}}%
}(\left\{ \lambda _{j}\right\} )$ for all possible configurations $\left\{
\lambda _{j}\right\} $. The bipartite entanglement entropy $S(\varepsilon )$%
\ for the eigenstate with energy $\varepsilon $ is then computed. The effect
of the Hilbert space fragmentation can be investigated by examining the
probability distribution of $S$, $P(S)$. To reduce computational cost, we
employ the Monte Carlo method by randomly selecting a portion of the
configurations $\left\{ \lambda _{j}\right\} $.

In order to reveal the effect of the resonance condition on the Hilbert
space fragmentation, we plot the functions $S(\varepsilon )\ $and\ $P(S)$
for three typical finite lattices in Fig. (\ref{fig2})(a), (b), and (c). We
find that the patterns not only depend on the parameters but also on the
geometry of the lattices. Fig. \ref{fig2} shows that the entanglement
entropy $S$ for a portion of eigenstates peaks at zero as expected when the
system is at resonance. In contrast, eigenstates with vanishing entanglement
entropy disappear when the system moves off the resonance.

\section{Quantum scars}

\label{Quantum scars}

Thus far, we have demonstrated that resonant side coupling induces Hilbert
space fragmentation in transverse-field Ising systems on arbitrary lattices.
In this section, we demonstrate that such systems host special nonthermal
eigenstates exhibiting periodic revivals. Inspired by recent experiments
with Rydberg atoms, where nonthermal periodic revival dynamics have been
observed for the initial Neel state \cite%
{bernien2017probing,bluvstein2021controlling}, we will examine the dynamics
for a simple initial state. We only focus on systems in which the Ising
model on lattice A is a uniform bipartite lattice. This allows us to
describe the scheme in a simple way. This method remains valid for complex
systems beyond this simple case.

To proceed, we consider the system defined on a uniform bipartite lattice,
where the Hamiltonian reads%
\begin{equation}
H=\sum_{\left\langle i,j\right\rangle }J\sigma _{a,i}^{z}\sigma
_{a,j}^{z}+\sum_{l}\sigma _{a,l}^{x}\left( g+\kappa \sigma _{b,l}^{x}\right)
.
\end{equation}%
where $i\in $odd and $j\in $even\ denote the two sublattices of the
bipartite lattice.\ Considering the invariant subspace in which the spins in
lattice B are in the state%
\begin{equation}
\left\vert \varphi \right\rangle _{\text{B}}=\prod_{j\in \text{odd}%
}\left\vert \leftarrow \right\rangle _{b,j}\prod_{l\in \text{even}%
}\left\vert \rightarrow \right\rangle _{b,l},  \label{psiB}
\end{equation}%
the eigenstates in such a subspace can be written in the form $\left\vert
\psi \right\rangle =\left\vert \varphi \right\rangle _{\text{A}}\left\vert
\varphi \right\rangle _{\text{B}}$, where $\left\vert \varphi \right\rangle
_{\text{A}}$ is the eigenstate of the Ising Hamiltonians of the form%
\begin{equation}
H_{\text{\textrm{A}}}=\sum_{\left\langle i,j\right\rangle }J\sigma
_{a,i}^{z}\sigma _{a,j}^{z}+\sum_{l}\left[ g+\kappa (-1)^{l}\right] \sigma
_{a,l}^{x},  \label{HA}
\end{equation}%
which represents an Ising model with a staggered transverse field. The
Hamiltonian $H_{\text{\textrm{A}}}$\ with nearest-neighbor (NN) coupling is
exactly solvable only on a 1D lattice. However, it becomes solvable for
lattices of any dimension under the resonance condition $g=\kappa $. In this
case, the subspace can be further decomposed into many subspaces. Here we
only consider the one that contains the state%
\begin{equation}
\left\vert \Downarrow \right\rangle =\prod_{\text{all }l}\left\vert
\downarrow \right\rangle _{a,l}\prod_{j\in \text{odd}}\left\vert \leftarrow
\right\rangle _{b,j}\prod_{l\in \text{even}}\left\vert \rightarrow
\right\rangle _{b,l}.  \label{FM}
\end{equation}%
In this subspace, the Hamiltonian $H_{\text{\textrm{A}}}$\ and the state $%
\left\vert \Downarrow \right\rangle $\ reduce to%
\begin{equation}
h_{\text{\textrm{A}}}=-ZJ\sum_{j\in \text{even}}\sigma
_{a,j}^{z}+g\sum_{j\in \text{even}}\left( \sigma _{a,l}^{+}+\sigma
_{a,l}^{-}\right) ,
\end{equation}%
and%
\begin{equation}
\left\vert \Downarrow \right\rangle =\prod_{l\in \text{even}}\left\vert
\downarrow \right\rangle _{a,l},
\end{equation}%
where $Z$ is the coordination number of the lattice. The reduced Hamiltonian 
$h_{\text{\textrm{A}}}$\ possesses an energy tower with spacing $2\sqrt{%
\left( ZJ\right) ^{2}+4g^{2}}$, which supports periodic dynamics.

The conclusion depends on the resonance condition. Accordingly, the
off-resonance parameter may cause the energy levels to deviate from the
energy tower, an effect that is usually non-negligible in practice. We focus
on how this off-resonance effect influences the energy tower. Our strategy
is to examine the dynamic response of the state $\left\vert \Downarrow
\right\rangle $ under a quenching process. Specifically, we numerically
compute the time evolution starting from the initial state $\left\vert \phi
\left( 0\right) \right\rangle =\left\vert \Downarrow \right\rangle $ under
the quenched Hamiltonian $H_{\text{\textrm{A}}}$ with $g\neq \kappa $. The
evolved state can be expressed as%
\begin{equation}
\left\vert \phi \left( t\right) \right\rangle =e^{-iH_{\text{\textrm{A}}%
}t}\left\vert \Downarrow \right\rangle ,
\end{equation}%
which is obtained by exact diagonalization for finite systems\ with several
typical geometry under the periodic boundary conditions. We employ the
fidelity, given by%
\begin{equation}
F(t)=\left\vert \left\langle \Downarrow \right\vert e^{-iH_{\text{\textrm{A}}%
}t}\left\vert \Downarrow \right\rangle \right\vert ^{2},  \label{fidelity}
\end{equation}%
to characterize the dynamic response\ induced by different values of $%
g-\kappa $. We plot $F(t)$ and $S(t)$ in Fig. \ref{fig3} as a function of ${t%
}$ for selected systems. The results show that the fidelity $F(t)$ exhibits
 perfect revivals and the entropy remains at zero when $g=\kappa $, as expected. However, in the off-resonance case, the revivals occur with fading amplitude.

Before ending this paper, we would like to address the comparison between
the PXP model and the present model. (i) Both models are quantum models
featuring QMBSs and only nearest-neighbor interactions are involved. (ii)
The blockade condition in the PXP model prevents adjacent spins from
simultaneously being in the spin-up state, described by the term $\left(
1-\sigma _{l-1}^{z}\right) \sigma _{l}^{x}\left( 1-\sigma _{l+1}^{z}\right) $%
. In our model, the blockade condition involves two sets of spins, described
by the term $\sigma _{a,l}^{x}\left( 1-\sigma _{b,l}^{x}\right) $. (iii)
This restricts the PXP model to a chain system, while our model can be on an
arbitrary lattice. (iv) The equivalent Hamiltonian for the quantum scars in
the PXP model corresponds to the graph of the Fibonacci chain, while in our
model it is a hypercube arising from the representation of a specific
angular momentum operator\cite{christandl2004perfect,zhang2012perfect}. In
this sense, while our model is simplified in terms of blockade condition, it
is more general in terms of lattice geometry.

\section{Conclusions}

\label{conclusions}

In summary, we have proposed a family of quantum spin systems to study
Hilbert space fragmentation and quantum scars. In contrast to the PXP model,
our system consists of two sets of quantum spins, A and B. Set A is the
parent system, which is an Ising model on arbitrary lattices with a
transverse field, while set B is the counterpart of set A, comprising free
spins that are coupled to set A. We have shown that the Hilbert space is
fragmented into an exponential number of decoupled sectors when the
transverse field and the side coupling strength are at resonance. To
demonstrate our scheme, several typical scar subspaces were studied
analytically. Numerical simulations of the probability distribution of
entanglement entropy for finite-size chains, square and triangular lattices
were performed using the Monte Carlo method. The results show that the
resonance condition is crucial for Hilbert space fragmentation and the
corresponding quantum scars. This finding reveals an alternative class of
quantum systems that exhibit subspaces that are immune to thermalization.

\acknowledgments This work was supported by the National Natural Science
Foundation of China (under Grant No. 12374461).

\section*{Data availability}

The data that support the findings of this article are openly
available \cite{ma_2026_19058423}.

\bibliography{HSF_reference.bib}

\end{document}